\documentclass[review]{elsarticle}

\usepackage{amssymb,amsfonts,amsthm}
\usepackage{amsmath,bm}
\usepackage{mathrsfs}
\usepackage{textcomp}
\usepackage{manyfoot}

\usepackage{graphicx}
\usepackage{booktabs}
\usepackage{multirow}
\usepackage{colortbl}
\usepackage{arydshln}
\usepackage[font=normalsize,labelfont=bf]{caption}
\usepackage{subcaption}

\usepackage{setspace}
\usepackage[right=1in, left=1in, top=1in, bottom=1in]{geometry} 
\usepackage{float} 
\usepackage{placeins}
\usepackage[utf8]{inputenc}
\usepackage{csquotes}

\usepackage{algorithm}
\usepackage{algpseudocode}
\usepackage{listings}

\biboptions{sort&compress,numbers}

\usepackage[colorlinks=true, linkcolor=blue, citecolor=blue, urlcolor=blue]{hyperref}

\hypersetup{
    pdfauthor={Reza Namakian, Fei Shuang, Thomas D Swinburne, Poulumi Dey, Ali Erdemir, Wei Gao},
    pdftitle={Reconciling Atomistic Predictions and Experimental Observations of stacking-faults in CoCrNi}
}

\graphicspath{{./figures/}}

\makeatletter
\def\ps@pprintTitle{%
 \let\@oddhead\@empty
 \let\@evenhead\@empty
 \def\@oddfoot{\reset@font\hfil\thepage\hfil}
 \let\@evenfoot\@oddfoot
}
\makeatother

\journal{Computational Materials Science}

\begin{document}

\begin{frontmatter}

\title{Finite Temperature Stacking Fault Stability in Random and Locally Ordered CoCrNi beyond the Harmonic Approximation}

\author[1]{Reza Namakian}
\author[2]{Fei Shuang}
\author[3,4]{Thomas D Swinburne}
\author[2]{Poulumi Dey}
\author[1,5]{Ali Erdemir}
\author[1,5]{Wei Gao\corref{correspondingauthor}}
\ead{wei.gao@tamu.edu}

\address[1]{J. Mike Walker $'$66 Department of Mechanical Engineering, Texas A\&M University, College Station, Texas 77843, United States}
\address[2]{Department of Materials Science and Engineering, Faculty of Mechanical Engineering, Delft University of Technology, Mekelweg 2, 2628 CD Delft, The Netherlands}
\address[3]{Aix-Marseille Universit´e, CNRS, CINaM UMR 7325, Campus de Luminy, 13288 Marseille, France}
\address[4]{Department of Mechanical Engineering,
University of Michigan, Ann Arbor, Michigan 48109, USA}
\address[5]{Department of Materials Science \& Engineering, Texas A\&M University, College Station, Texas 77843, United States}
\cortext[correspondingauthor]{Corresponding author}

\begin{abstract}

Previous density functional theory (DFT) calculations for random solid solution (RSS) CoCrNi predict negative intrinsic stacking-fault energy (ISFE) at 0 K, contrary to experimental observations of finite stacking-fault widths. Two explanations have been proposed: finite-temperature stabilization of the RSS state, suggested by harmonic approximations showing increasing ISFE with temperature, and local chemical order (LCO), which shifts the ISFE to positive values at 0 K. Here, we compute temperature-dependent generalized stacking-fault free energies for RSS and LCO CoCrNi using a near-quantum-accuracy machine learning interatomic potential and the fully anharmonic projected average force integrator. Unlike harmonic approximations, our anharmonic calculations show that the RSS ISFE decreases with temperature and remains negative, indicating that RSS stacking faults are not thermally stabilized at elevated temperatures. By contrast, LCO maintains positive ISFE over 0–1000 K.  Molecular dynamics simulations further confirm unbounded dislocation dissociation in RSS CoCrNi but finite stacking-fault widths in the LCO state.

\end{abstract}

\end{frontmatter}

Multi-principal element alloys (MPEAs) have established themselves as a premier class of structural materials, distinguished by their exceptional thermal, mechanical and environmental performance. Within this family, face-centered cubic (FCC) systems—specifically the equiatomic CoCrNi medium-entropy alloy—have been intensively studied for their remarkable synergy of strength, ductility, and fracture toughness, a balance that notably persists even at cryogenic temperatures \cite{liu2022exceptional}.
In these FCC MPEAs, plastic deformation is primarily mediated by dislocation slip on \textnormal{\{111\}} planes. The generalized stacking-fault energy (GSFE) curve maps the energetic landscape of this slip process \cite{namakian2023temperature}, yielding key metrics such as the intrinsic and unstable stacking-fault energies (ISFE and USFE, respectively). These parameters are critical for predicting diverse phenomena, from dislocation dynamics and deformation twinning to the FCC-to-HCP martensitic phase transformation \cite{zhu2023effects}. An important feature of MPEAs is the strong spatial variation in ISFE and USFE arising from local compositional fluctuations. This chemical heterogeneity creates a rugged GSFE landscape, causing dislocations to advance through stick-slip motion of nanoscale segments, a process often described as ``nanoscale segment detrapping'' \cite{ma2024chemical, utt2022origin, li2023fluctuations}.  Thus, the GSFE curve serves as a useful descriptor for mechanical behavior because it captures the local chemical and elastic interactions that govern dislocation motion \cite{shuang2025standard}.

Notably, a discrepancy exists between previous density functional theory (DFT) predictions and experimental observations regarding stacking-fault stability in Cantor-family alloys. Finite stacking-fault width (SFW) has been observed experimentally \cite{laplanche2017reasons, liu2018stacking, li2022tensile}. However, DFT calculations for alloys modeled as random solid solutions (RSS) predict negative mean ISFE values at 0 K \cite{zhang2013first, niu2018magnetically,zhao2017stacking}, which suggests continuous separation of the partial dislocations and an indefinitely expanding stacking-fault.

There have been two explanations proposed to resolve this discrepancy. The first attributes the discrepancy to finite-temperature effects. DFT-based free energy calculations using harmonic or quasi-harmonic approximations for the RSS state suggest that the ISFE increases with temperature and can become positive at elevated temperatures \cite{BARUFFI2023115536, zhao2017stacking, zhang2017dislocation, niu2018magnetically}. The second explanation involves local chemical order (LCO), which is now recognized as ubiquitous in MPEAs \cite{ma2025compositional, han2024ubiquitous, yang2024rejuvenation}. In this view, LCO provides a possible thermodynamic mechanism for stabilizing finite stacking-faults. Both DFT simulations \cite{li2019strengthening, zhu2023effects} and empirical embedded-atom method (EAM) calculations \cite{Ding2018} have shown that LCO can shift the mean ISFE from negative to positive values at 0 K.

This study is motivated by two key gaps in the existing explanations. First, previous finite-temperature ISFE calculations for RSS structures have neglected explicit anharmonic effects. This approximation can be problematic for stacking-fault thermodynamics. For example, Xu et al. \cite{XU2024115934} recently showed that for Ni\textsubscript{3}Al the quasi-harmonic contribution alone predicts an increasing complex stacking-fault energy with temperature, whereas the inclusion of explicit anharmonic vibrations reverses this trend and leads to a decreasing complex stacking-fault Gibbs energy at elevated temperatures. Second, although LCO has been shown to shift the mean ISFE from negative to positive values at 0 K, its effect on the ISFE at finite temperatures remains unexplored. Therefore, it remains unclear whether the positive ISFE induced by LCO persists at elevated temperatures.

In this work, we investigate the temperature dependence of the GSFE in CoCrNi with both RSS and LCO configurations, explicitly accounting for anharmonic vibrational effects using the fully-anharmonic projected average force integrator (PAFI) \cite{swinburne2018unsupervised, namakian2023temperature, namakian2024temperature}. To efficiently sample diverse local atomic environments, we employ a near-quantum-accuracy neural network potential (NNP) \cite{du2022chemical}. We also corroborate our findings by explicitly simulating edge dislocation structures in large-scale models for both chemical states. All atomistic simulations were performed using the Large-Scale Atomic/Molecular Massively Parallel Simulator (LAMMPS) \cite{LAMMPS}. Initial atomic structures were constructed using the Atomsk package \cite{Hirel2015Atomsk}. Post-processing and defect analysis were conducted using OVITO Pro \cite{ovito}, using the Interval Common Neighbor Analysis (ICNA) technique \cite{Larsen2020Revisiting} to identify local structural variations.

\begin{figure}[t!]
    \centering
    \includegraphics[width=\textwidth]{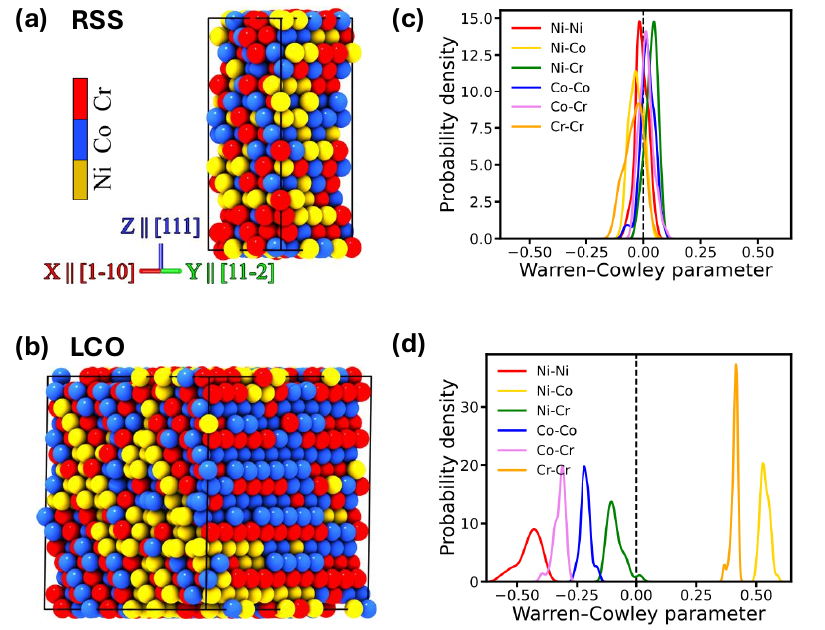}
    \caption{\textbf{(a, b)} Representative atomic configurations of \textbf{(a)} the 450-atom RSS supercell and \textbf{(b)} the 2,520-atom LCO supercell containing nanometer-sized domains. \textbf{(c, d)} Probability density distributions of the first-nearest-neighbor Warren-Cowley parameters calculated across the ensemble of 50 RSS and 50 LCO configurations, respectively.}
    \label{fig:rss_lco_wc}
\end{figure}

To rigorously account for the statistical fluctuations inherent in the RSS configurations, we analyzed an ensemble of 50 independent atomic configurations, each containing 450 atoms. These RSS supercells were constructed by randomly distributing equimolar fractions of Co, Cr, and Ni. A representative configuration is shown in Fig.~\ref{fig:rss_lco_wc}a.
For LCO configurations, we prepared an ensemble of 50 independent 2,520-atom supercells with approximate dimensions of $L_X \approx L_Y \approx L_Z \approx 3$ nm, designed to capture chemically ordered domains with characteristic length scales extending beyond the nearest-neighbor shells. A representative structure is displayed in Fig.~\ref{fig:rss_lco_wc}b. These LCO states were generated via a hybrid molecular dynamics/Monte Carlo (MD/MC) protocol. The MC stage utilized canonical atom swaps to access low-energy, chemically ordered configurations. Subsequently, each system was equilibrated for 200 ps in the isothermal-isobaric (NPT) ensemble at 300 K and 0 bar, utilizing a Nosé-Hoover thermostat and barostat with a 1 fs timestep.

The chemical differences between the RSS and LCO configurations were quantified using Warren–Cowley (WC) parameters calculated over the corresponding 50-configuration ensembles, as shown in Figs.~\ref{fig:rss_lco_wc}c and~\ref{fig:rss_lco_wc}d, respectively. For the RSS state, the WC distributions for all pair types are centered near zero, confirming an approximately random distribution of species. By contrast, the LCO state exhibits pair-specific WC distributions that are strongly shifted away from zero, indicating pronounced non-random chemical environments with both preferred and avoided elemental pairings. Notably, previous NNP-based MD/MC studies \cite{du2022chemical} using similarly sized 2,916-atom supercells have shown that annealed CoCrNi can develop LCO domains with characteristic sizes of 2–3 nm at temperatures below 700 K.

\begin{figure}[b!]
    \centering
    \includegraphics[width=\textwidth]{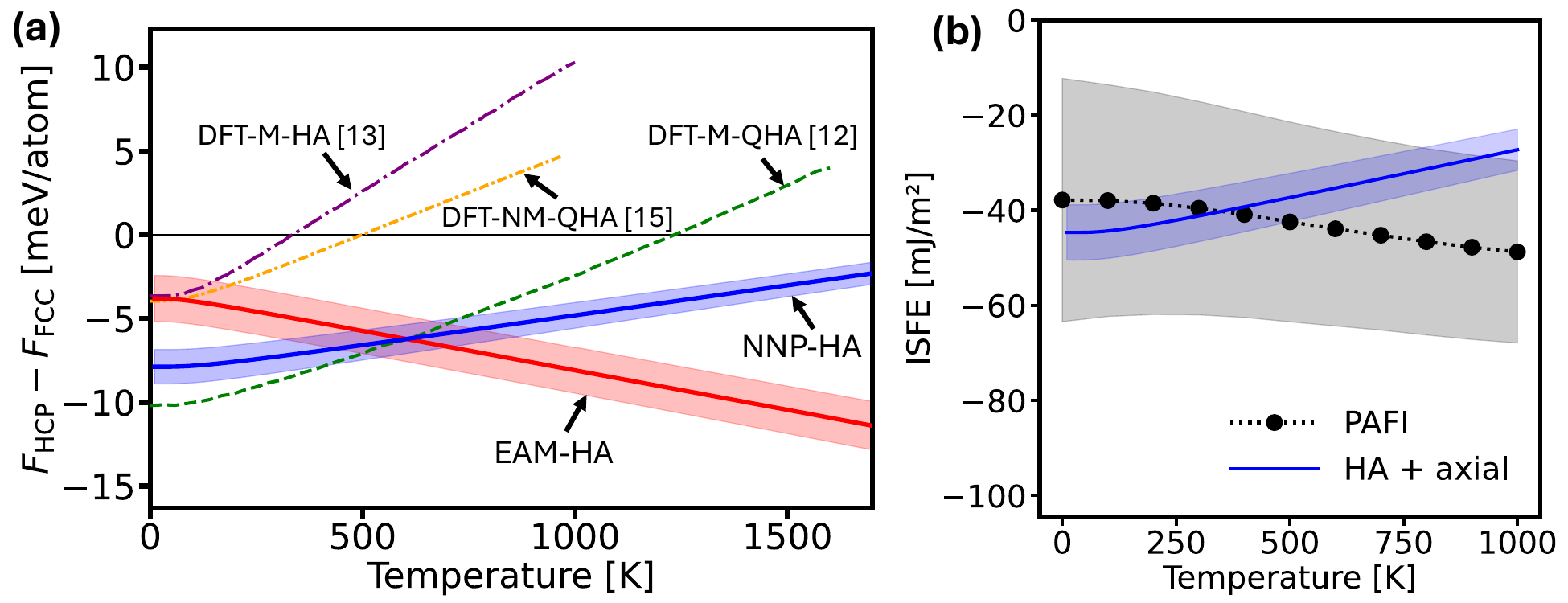}
    \caption{\textbf{(a)} HCP–FCC free energy difference. Our EAM and NNP results obtained within the harmonic approximation are compared with literature data using different magnetic and vibrational approximations: magnetic (M) or nonmagnetic (NM) calculations, and harmonic (HA) or quasi-harmonic (QHA) approximations. The shaded blue and red regions indicate the standard deviation across an ensemble of 50 unique random configurations. \textbf{(b)} ISFE for the RSS state. The solid blue line and accompanying shaded blue region represent the mean and standard deviation, respectively, calculated from the HCP-FCC free energy difference via first-order axial interaction model. For comparison, results from the fully anharmonic PAFI method are shown by the dashed black line with circular markers (mean values) and the transparent gray shaded region (standard deviation).}
    \label{fig:potential_validation_pafi_rss}
\end{figure}

To validate the selected interatomic potential \cite{du2022chemical} in capturing stacking-fault energetics, we first evaluated the temperature-dependent free-energy difference between the FCC and HCP phases in RSS CoCrNi and compared the results with available DFT data from the literature \cite{niu2018magnetically,zhao2017stacking,zhang2017dislocation}. As shown in Fig.~\ref{fig:potential_validation_pafi_rss}a, the reported DFT values vary depending on the magnetic treatment and the vibrational approximations used (harmonic or quasi-harmonic approximations). Nevertheless, all DFT results exhibit a consistent trend: the free energy difference increases with temperature. Our NNP calculations within the harmonic approximation reproduce the same trend. The standard deviation of the NNP curve represents the  statistical fluctuations among different RSS configurations.
By contrast, calculations using the widely adopted embedded-atom method (EAM) potential \cite{li2019strengthening} predict the opposite temperature dependence. This discrepancy may arise from the known limitations of empirical potentials constructed via the force-matching method, which generally do not achieve the accuracy of well-trained machine learning interatomic potentials. For the CoCrNi EAM potential specifically, the mean energy deviation from DFT benchmarks was reported to be approximately 20 meV/atom \cite{li2019strengthening}. Therefore, despite its higher computational cost, the NNP was selected for the present study.

The temperature-dependent GSFE curves corresponding to intrinsic stacking-fault formation on the $\{111\}\langle 110\rangle$ plane were computed via a multi-stage procedure. First, the supercells with fully periodic boundary conditions were relaxed to a zero-stress state using the ABC-FIRE algorithm \cite{ECHEVERRIRESTREPO2023111978}. To accommodate lattice relaxation normal to the slip plane during faulting, the boundary conditions along the $Z$-axis were subsequently modified to introduce free surfaces. The minimum energy path (MEP) was then determined using the nudged elastic band (NEB) method \cite{HenkelmanA, HenkelmanB, Ghasemi_2019}. This 0 K MEP served as the input for the finite-temperature PAFI calculations. By integrating the thermally averaged mean force along this path, PAFI determined the corresponding minimum free-energy path while accounting for anharmonic vibrational contributions.  The final temperature-dependent GSFE curves were obtained by normalizing the integrated free energy by the fault-plane area, which was adjusted at each temperature to capture thermal expansion effects. To prevent unphysical relaxation along the slip direction during these calculations, a small subset of atoms on the fault-plane was constrained following the protocol established in \cite{namakian2023temperature}.

The temperature-dependent ISFE values extracted from the GSFE curves of the RSS configurations are presented in Fig.~\ref{fig:potential_validation_pafi_rss}b. For comparison, we also converted the HCP-FCC free energy difference, $F_{\mathrm{HCP}}(T)-F_{\mathrm{FCC}}(T)$, obtained from the harmonic approximation and shown in Fig.~\ref{fig:potential_validation_pafi_rss}a, into an approximated ISFE using the first-order axial interaction model \cite{denteneer1991energetics, denteneer1987stacking}, $\gamma_{\scriptscriptstyle \mathrm{ISF}}(T) \approx 2\left[F_{\mathrm{HCP}}(T)-F_{\mathrm{FCC}}(T)\right]/{A}$, where $A$ is the stacking-fault area. The two approaches predict different temperature trends. The harmonic model suggests that the ISFE increases with temperature, which has led previous studies to propose that the finite-temperature effect may stabilize stacking-faults \cite{BARUFFI2023115536}. By contrast, the PAFI results, which explicitly account for anharmonic vibrational effects, show that the ISFE decreases with increasing temperature. Anharmonic vibrational effects arise from the asymmetry of atomic vibrations caused by short-range repulsive interactions and can be particularly pronounced near fault-planes \cite{XU2024115934}. Our results suggest that stacking-faults in RSS CoCrNi are not thermally stabilized at elevated temperatures. It is noted that the difference between the PAFI and harmonic results at 0 K comes from the axial-interaction approximation used in the harmonic estimate.

Fig.~\ref{fig:potential_validation_pafi_rss}b shows that the ISFE variation across the 50 RSS configurations is significantly broader than that predicted by the harmonic axial interaction model. This difference is expected because the axial interaction model reduces the ISFE to a bulk free-energy difference and therefore neglects the explicit interfacial environment, local chemical fluctuations, and localized structural relaxation associated with the stacking-fault. By contrast, the PAFI calculations are performed directly on faulted configurations, allowing these local effects to be captured. Quantum thermal statistics were excluded from the present free-energy calculations, which may affect the low-temperature behavior. Nevertheless, the predicted decreasing trend and the conclusion are expected to remain valid at near room temperature and above, where quantum effects are negligible.

\begin{figure}[H]
    \centering
    \includegraphics[width=0.95\textwidth]{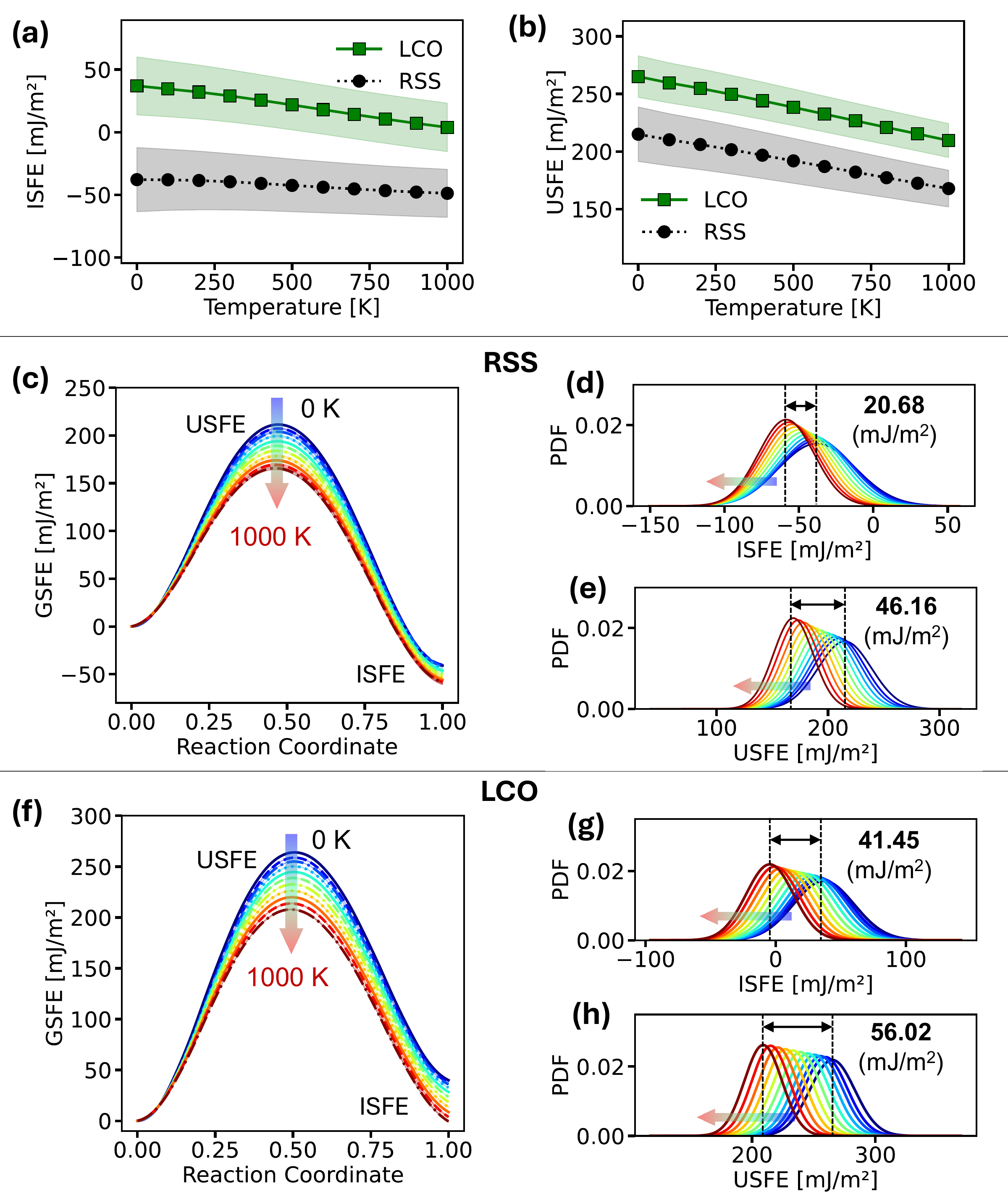}
    \caption{\textbf{(a)} ISFE and \textbf{(b)} USFE calculated via the PAFI method for an ensemble of 50 configurations in both RSS and LCO states. Solid lines represent mean values, while shaded regions denote the standard deviation. \textbf{(c, f)} GSFE curves for two representative configurations of the RSS and LCO states, respectively. \textbf{(d, e)} and \textbf{(g, h)} Probability density functions, fitted with Gaussian distributions, illustrating the statistical spread and thermal shift of the ISFE and USFE across the 50-configuration ensembles for the RSS and LCO states, respectively. Vertical dashed lines indicate the mean values at 0~K and 1000~K, with the labeled double-headed arrows quantifying the total thermal shifts.}
    \label{fig:gsffe_summary}
\end{figure}

We next evaluate how LCO alters the ISFE as a function of temperature. Fig.~\ref{fig:gsffe_summary}a reveals that the presence of LCO domains effectively stabilizes the stacking-fault by elevating the ISFE to notably high positive values relative to the RSS state. Specifically, the LCO CoCrNi system exhibits a mean ISFE of $\approx$ 38 mJ/m$^2$ at 0 K, distinguishing it sharply from the RSS baseline. While the LCO ISFE decreases monotonically with increasing temperature, driven by anharmonic atomic vibrations similar to behavior observed in elemental FCC metals like Cu \cite{namakian2023temperature}, it remains positive across nearly the entire temperature range, only approaching zero near 1000 K. This positive ISFE introduces a thermodynamic penalty for expanding the stacking-fault ribbon, providing the restoring force needed to balance the repulsive elastic interaction between partial dislocations. As a result, the LCO state supports a finite partial dislocation separation, whereas the negative-ISFE RSS state favors continuous expansion of the stacking-fault ribbon.

The temperature dependence of the USFE can also be extracted from the calculated GSFE curves. As shown in Fig.~\ref{fig:gsffe_summary}b, the USFE of the LCO state remains higher than that of the RSS state across the entire temperature range. This indicates that LCO increases the energy barrier to partial dislocation motion and therefore enhances the resistance to slip.
The GSFE results further show that both the USFE and ISFE exhibit comparable levels of statistical variation arising from differences in local chemical environments. This observation is consistent with previous work showing that the spread of stacking-fault energies caused by local chemical environment fluctuations remains nearly invariant across structures with different degrees of chemical short-range order \cite{Ding2018}. Our results further demonstrate that this statistical variation is also largely preserved with increasing temperature, suggesting that finite-temperature effect does not alter the spatial energy fluctuations characteristic of chemically complex environments in CoCrNi.

From the GSFE curves, we note that chemical disorder alters the relative temperature dependence of the USFE and ISFE. Unlike pure FCC metals such as Cu, where the USFE and ISFE exhibit similar temperature-dependent shifts and thus a relatively coupled thermal response along the GSFE path \cite{namakian2023temperature}, chemically disordered CoCrNi shows a weakened coupling between these two fault energies. Figs.~\ref{fig:gsffe_summary}c and~\ref{fig:gsffe_summary}f show the GSFE curves for representative RSS and LCO configurations, respectively. For both cases, the USFE and ISFE evolve differently with temperature, with the USFE exhibiting a stronger thermal decrease than the ISFE. This effect is most pronounced in the RSS state, where the mean USFE shifts by approximately 2.2 times more than the mean ISFE over the studied temperature range (shown in Figs.~\ref{fig:gsffe_summary}d and~\ref{fig:gsffe_summary}e). In the LCO state, this ratio decreases to approximately 1.35 (shown in Figs.~\ref{fig:gsffe_summary}g and~\ref{fig:gsffe_summary}h), indicating that local chemical order partially restores a more coupled thermal response between the unstable and intrinsic fault configurations. The stronger temperature sensitivity of the USFE in the RSS state likely arises because the USFE corresponds to a transition-state configuration along the shear path; in the RSS case, this structural distortion is superimposed on a broad distribution of local Co–Cr–Ni bonding environments, making the unstable fault configuration more sensitive to thermal vibrations.

\begin{figure}[t!]
    \centering
    \includegraphics[width=\textwidth]{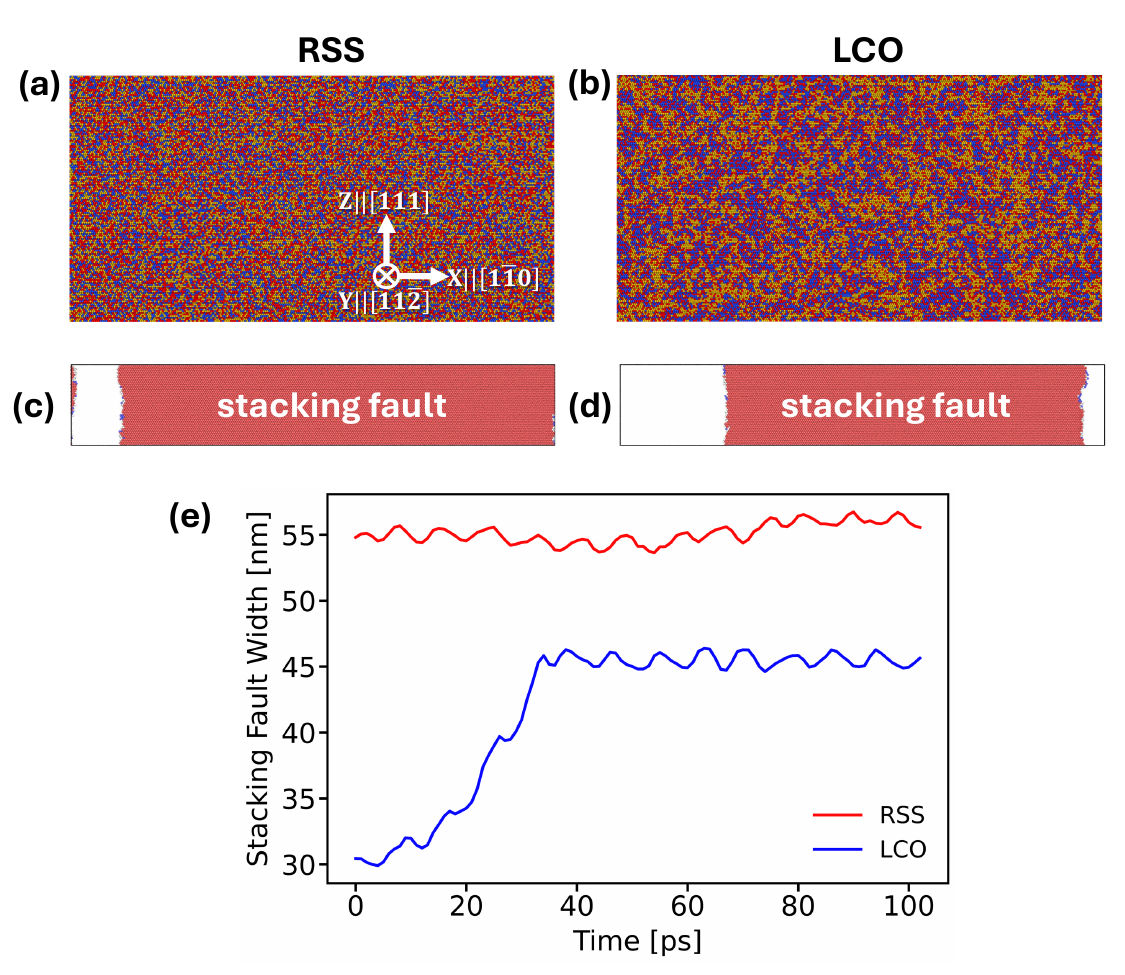}
    \caption{Large-scale edge dislocation simulations in CoCrNi. \textbf{(a, b)} 1.67-million atom supercells ($9.9 \times 60.2 \times 30.4$ nm) showing the initial RSS and LCO configurations embedding an edge dislocation. \textbf{(c, d)} Equilibrated edge dislocation core structures at 300 K for RSS and LCO states, respectively. Atoms are identified via ICNA and colored by local structure: HCP (red), BCC (blue), and Other (gray); FCC atoms are hidden for clarity. \textbf{(c)} illustrates complete dissociation across the supercell in the RSS state, while \textbf{(d)} shows a stabilized, finite SFW in the LCO state. \textbf{(e)} Time evolution of the SFW during 100 ps of MD equilibration at 300 K for both the RSS and LCO configurations.}
    \label{fig:Dislocation_LCO_and_RSS}
\end{figure}

To examine stacking-fault stability, we also performed MD simulations of edge dislocations in both the RSS and LCO states. These simulations used large-scale supercells containing approximately 1.67 million atoms, with dimensions of $L_X \approx 9.9$ nm, $L_Y \approx 60.2$ nm, and $L_Z \approx 30.4$ nm, as shown in Fig.~\ref{fig:Dislocation_LCO_and_RSS}. Notably, the large system sizes were chosen to enable more robust sampling of the local chemical environments surrounding the dislocation core. Because hybrid MD/MC simulations were computationally prohibitive at this scale, the LCO dislocation supercell was generated using the reverse Monte Carlo (RMC) algorithm \cite{doi:10.1073/pnas.2322962121}. The RMC procedure matched target WC parameters derived from 50 smaller, equilibrated LCO configurations shown in Fig.~\ref{fig:rss_lco_wc}d. The resulting structure was then energy-minimized and equilibrated for 100 ps at 300 K and zero pressure in the NPT ensemble. An edge dislocation was introduced into each supercell, with free surfaces applied along the $Z$-direction to reduce long-range stress fields normal to the slip plane.

In the RSS supercell (Fig.~\ref{fig:Dislocation_LCO_and_RSS}c), the dislocation exhibits unbounded dissociation immediately upon energy minimization and remains unbounded following equilibration at 300 K. This behavior is quantitatively captured by the MD time evolution of the stacking-fault width in Fig.~\ref{fig:Dislocation_LCO_and_RSS}e, where the fault expands to the periodic boundaries. At this limit, the two partials remain separated solely by their mutual elastic repulsion, reflecting the thermodynamic instability associated with the negative ISFE values shown in Fig.~\ref{fig:gsffe_summary}a. By contrast, the LCO state confines the dislocation core. As shown in Figs.~\ref{fig:Dislocation_LCO_and_RSS}d and~\ref{fig:Dislocation_LCO_and_RSS}e, the stacking-fault width increases during the initial $\sim$40 ps of equilibration at 300 K before reaching a finite, stable separation for the remainder of the simulation. This behavior indicates that LCO provides a restoring force against further fault expansion and stabilizes a finite stacking-fault width at finite temperatures, consistent with our free-energy calculations.

In summary, we investigated the finite-temperature stacking-fault stability of CoCrNi in both RSS and LCO states using fully anharmonic free energy calculations and large-scale MD simulations. In contrast to harmonic approximations, which predict that the ISFE increases with temperature, the direct PAFI calculations show that the ISFE of RSS CoCrNi decreases with increasing temperature and remains negative. This result indicates that finite-temperature effects do not stabilize stacking-faults in the RSS state. By contrast, LCO shifts the ISFE to positive values, providing the thermodynamic restoring force required to stabilize finite partial dislocation separations. Large-scale MD dislocation simulations further corroborate this result: the RSS state exhibits unbounded dissociation, whereas the LCO state maintains a finite stacking-fault width at 300 K. Together, these results show that local chemical order, rather than thermal stabilization of the random alloy, provides the more plausible origin of finite stacking-fault widths in CoCrNi at finite temperatures. Finally, the GSFE curves reveal that chemical disorder alters the relative thermal responses of the USFE and ISFE: in the RSS state, the USFE decreases much more sharply with temperature than the ISFE, whereas LCO partially restores a more coupled temperature dependence between these two fault energies.

\section*{Data Availability}
The data that support the findings of this study are available upon reasonable request.

\section*{CRediT authorship contribution statement}
\textbf{Reza Namakian}: Methodology, Software, Investigation, Writing - original draft, Writing - review \& editing. \textbf{Fei Shuang}: Methodology, Software, Investigation, Writing - review \& editing. \textbf{Thomas D Swinburne}: Methodology, Software, Investigation, Writing - review \& editing. \textbf{Poulumi Dey}: Investigation, Writing - review \& editing. \textbf{Ali Erdemir}: Reviewing \& editing.  \textbf{Wei Gao}: Conceptualization, Methodology, Software, Investigation, Writing - original draft, Writing - review \& editing, Supervision, Funding acquisition.

\section*{Acknowledgments}
W.G. gratefully acknowledges financial support of this work by the National Science Foundation through Grant No. CMMI-2305529. The authors acknowledge the Texas Advanced Computing Center (TACC) at the University of Texas at Austin and Texas A\&M High Performance Research Computing for providing HPC resources that have contributed to the research results reported within this paper. This work also used DeltaAI at NCSA - University of Illinois through allocation MCH250077 from the Advanced Cyberinfrastructure Coordination Ecosystem: Services \& Support (ACCESS) program, which is supported by U.S. National Science Foundation grants \#2138259, \#2138286, \#2138307, \#2137603, and \#2138296 \cite{Boerner_2023}.

\bibliographystyle{elsarticle-num} 
\bibliography{ref}

@article{liu2022exceptional,
  author = {Dong Liu and Qin Yu and Saurabh Kabra and Ming Jiang and Paul Forna-Kreutzer and Ruopeng Zhang and Madelyn Payne and Flynn Walsh and Bernd Gludovatz and Mark Asta and Andrew M. Minor and Easo P. George and Robert O. Ritchie},
  title = {Exceptional fracture toughness of CrCoNi-based medium- and high-entropy alloys at 20 kelvin},
  journal = {Science},
  volume = {378},
  number = {6623},
  pages = {978-983},
  year = {2022},
  doi = {10.1126/science.abp8070},
  url = {https://www.science.org/doi/abs/10.1126/science.abp8070},
  eprint = {https://www.science.org/doi/pdf/10.1126/science.abp8070}
}

@article{ma2024chemical,
  title = {Chemical inhomogeneities in high-entropy alloys help mitigate the strength-ductility trade-off},
  author = {Evan Ma and Chang Liu},
  journal = {Progress in Materials Science},
  volume = {143},
  pages = {101252},
  year = {2024},
  issn = {0079-6425},
  doi = {10.1016/j.pmatsci.2024.101252},
  url = {https://www.sciencedirect.com/science/article/pii/S0079642524000215}
}

@article{ma2025compositional,
  title = {Compositional fluctuation and local chemical ordering in multi-principal element alloys},
  author = {Evan Ma and Jun Ding},
  journal = {Journal of Materials Science \& Technology},
  volume = {220},
  pages = {233-244},
  year = {2025},
  issn = {1005-0302},
  doi = {10.1016/j.jmst.2024.09.008},
  url = {https://www.sciencedirect.com/science/article/pii/S1005030224009113}
}

@article{namakian2023temperature,
  title = {Temperature dependent stacking fault free energy profiles and partial dislocation separation in FCC Cu},
  author = {Reza Namakian and Dorel Moldovan and Thomas D. Swinburne},
  journal = {Computational Materials Science},
  volume = {218},
  pages = {111971},
  year = {2023},
  issn = {0927-0256},
  doi = {10.1016/j.commatsci.2022.111971},
  url = {https://www.sciencedirect.com/science/article/pii/S0927025622006826}
}

@article{shuang2025standard,
  title = {Standard deviation in maximum restoring force controls the intrinsic strength of face-centered cubic multi-principal element alloys},
  author = {Fei Shuang and Luca Laurenti and Poulumi Dey},
  journal = {Acta Materialia},
  volume = {282},
  pages = {120508},
  year = {2025},
  issn = {1359-6454},
  doi = {10.1016/j.actamat.2024.120508},
  url = {https://www.sciencedirect.com/science/article/pii/S1359645424008577}
}

@article{li2023fluctuations,
  title = {Fluctuations in local shear-fault energy produce unique and dominating strengthening in metastable complex concentrated alloys},
  author = {Wei Li and Shuang Lyu and Yue Chen and Alfonso H. W. Ngan},
  journal = {Proceedings of the National Academy of Sciences},
  volume = {120},
  number = {12},
  pages = {e2209188120},
  year = {2023},
  doi = {10.1073/pnas.2209188120},
  url = {https://www.pnas.org/doi/abs/10.1073/pnas.2209188120},
  eprint = {https://www.pnas.org/doi/pdf/10.1073/pnas.2209188120}
}

@article{utt2022origin,
  title = {The origin of jerky dislocation motion in high-entropy alloys},
  author = {Daniel Utt and Subin Lee and Yaolong Xing and Hyejin Jeong and Alexander Stukowski and Sang Ho Oh and Gerhard Dehm and Karsten Albe},
  journal = {Nature Communications},
  volume = {13},
  number = {1},
  pages = {4777},
  year = {2022},
  doi = {10.1038/s41467-022-32134-1},
  url = {https://doi.org/10.1038/s41467-022-32134-1}
}

@article{zhu2023effects,
  title = {Effects of short range ordering on the generalized stacking fault energy and deformation mechanisms in FCC multiprincipal element alloys},
  author = {Lingyu Zhu and Zhaoxuan Wu},
  journal = {Acta Materialia},
  volume = {259},
  pages = {119230},
  year = {2023},
  issn = {1359-6454},
  doi = {10.1016/j.actamat.2023.119230},
  url = {https://www.sciencedirect.com/science/article/pii/S1359645423005608}
}

@article{laplanche2017reasons,
  title = {Reasons for the superior mechanical properties of medium-entropy CrCoNi compared to high-entropy CrMnFeCoNi},
  author = {G. Laplanche and A. Kostka and C. Reinhart and J. Hunfeld and G. Eggeler and E.P. George},
  journal = {Acta Materialia},
  volume = {128},
  pages = {292-303},
  year = {2017},
  issn = {1359-6454},
  doi = {10.1016/j.actamat.2017.02.036},
  url = {https://www.sciencedirect.com/science/article/pii/S135964541730126X}
}

@article{li2022tensile,
  title = {Tensile and compressive plastic deformation behavior of medium-entropy {Cr-Co-Ni} single crystals from cryogenic to elevated temperatures},
  author = {Le Li and Zhenghao Chen and Shogo Kuroiwa and Mitsuhiro Ito and Kyosuke Kishida and Haruyuki Inui and Easo P. George},
  journal = {International Journal of Plasticity},
  volume = {148},
  pages = {103144},
  year = {2022},
  issn = {0749-6419},
  doi = {10.1016/j.ijplas.2021.103144},
  url = {https://www.sciencedirect.com/science/article/pii/S0749641921002126}
}

@article{zhao2017stacking,
  title = {Stacking fault energies of face-centered cubic concentrated solid solution alloys},
  author = {Shijun Zhao and G. Malcolm Stocks and Yanwen Zhang},
  journal = {Acta Materialia},
  volume = {134},
  pages = {334-345},
  year = {2017},
  issn = {1359-6454},
  doi = {10.1016/j.actamat.2017.05.001},
  url = {https://www.sciencedirect.com/science/article/pii/S1359645417303671}
}

@article{zhang2017dislocation,
  title = {Dislocation mechanisms and {3D} twin architectures generate exceptional strength-ductility-toughness combination in {CrCoNi} medium-entropy alloy},
  author = {Zijiao Zhang and Hongwei Sheng and Zhangjie Wang and Bernd Gludovatz and Ze Zhang and Easo P. George and Qian Yu and Scott X. Mao and Robert O. Ritchie},
  journal = {Nature Communications},
  volume = {8},
  number = {1},
  pages = {14390},
  year = {2017},
  doi = {10.1038/ncomms14390},
  url = {https://doi.org/10.1038/ncomms14390}
}

@article{niu2018magnetically,
  title = {Magnetically-driven phase transformation strengthening in high entropy alloys},
  author = {Changning Niu and Carlyn R. LaRosa and Jiashi Miao and Michael J. Mills and Maryam Ghazisaeidi},
  journal = {Nature Communications},
  volume = {9},
  number = {1},
  pages = {1363},
  year = {2018},
  doi = {10.1038/s41467-018-03846-0},
  url = {https://doi.org/10.1038/s41467-018-03846-0}
}

@article{li2019strengthening,
  title = {Strengthening in multi-principal element alloys with local-chemical-order roughened dislocation pathways},
  author = {Qing-Jie Li and Howard Sheng and Evan Ma},
  journal = {Nature Communications},
  volume = {10},
  number = {1},
  pages = {3563},
  year = {2019},
  doi = {10.1038/s41467-019-11464-7},
  url = {https://doi.org/10.1038/s41467-019-11464-7}
}

@article{yang2024rejuvenation,
  title = {Rejuvenation as the origin of planar defects in the CrCoNi medium entropy alloy},
  author = {Yang Yang and Sheng Yin and Qin Yu and Yingxin Zhu and Jun Ding and Ruopeng Zhang and Colin Ophus and Mark Asta and Robert O. Ritchie and Andrew M. Minor},
  journal = {Nature Communications},
  volume = {15},
  number = {1},
  pages = {1402},
  year = {2024},
  doi = {10.1038/s41467-024-45696-z},
  url = {https://doi.org/10.1038/s41467-024-45696-z}
}

@article{han2024ubiquitous,
  title = {Ubiquitous short-range order in multi-principal element alloys},
  author = {Ying Han and Hangman Chen and Yongwen Sun and Jian Liu and Shaolou Wei and Bijun Xie and Zhiyu Zhang and Yingxin Zhu and Meng Li and Judith Yang and Wen Chen and Penghui Cao and Yang Yang},
  journal = {Nature Communications},
  volume = {15},
  number = {1},
  pages = {6486},
  year = {2024},
  doi = {10.1038/s41467-024-49606-1},
  url = {https://doi.org/10.1038/s41467-024-49606-1}
}

@article{du2022chemical,
  title = {Chemical domain structure and its formation kinetics in CrCoNi medium-entropy alloy},
  author = {Jun-Ping Du and Peijun Yu and Shuhei Shinzato and Fan-Shun Meng and Yuji Sato and Yangen Li and Yiwen Fan and Shigenobu Ogata},
  journal = {Acta Materialia},
  volume = {240},
  pages = {118314},
  year = {2022},
  issn = {1359-6454},
  doi = {10.1016/j.actamat.2022.118314},
  url = {https://www.sciencedirect.com/science/article/pii/S1359645422006930}
}

@article{swinburne2018unsupervised,
  title = {Unsupervised Calculation of Free Energy Barriers in Large Crystalline Systems},
  author = {Thomas D. Swinburne and Mihai-Cosmin Marinica},
  journal = {Physical Review Letters},
  volume = {120},
  number = {13},
  pages = {135503},
  year = {2018},
  month = {Mar},
  publisher = {American Physical Society},
  doi = {10.1103/PhysRevLett.120.135503},
  url = {https://link.aps.org/doi/10.1103/PhysRevLett.120.135503}
}

@article{namakian2024temperature,
  title = {Temperature dependence of generalized stacking fault free energy profiles and dissociation mechanisms of slip systems in Mg},
  author = {Reza Namakian and Dorel Moldovan and Thomas D. Swinburne},
  journal = {Computational Materials Science},
  volume = {231},
  pages = {112569},
  year = {2024},
  issn = {0927-0256},
  doi = {10.1016/j.commatsci.2023.112569},
  url = {https://www.sciencedirect.com/science/article/pii/S0927025623005633}
}

@Article{LAMMPS,
  author = "A. P. Thompson and H. M. Aktulga and R. Berger and 
     D. S. Bolintineanu and W. M. Brown and P. S. Crozier and
     P. J. in 't Veld and A. Kohlmeyer and S. G. Moore and T. D. Nguyen and
     R. Shan and M. J. Stevens and J. Tranchida and C. Trott and S. J. Plimpton",
  title = "{LAMMPS} - a flexible simulation tool for
     particle-based materials modeling at the 
     atomic, meso, and continuum scales",
  journal = "Comp. Phys. Comm.",
  volume =  "271",
  pages =   "108171",
  year =    "2022",
  doi = "10.1016/j.cpc.2021.108171"
}

@article{ ovito,
Author = {Stukowski, Alexander},
Title = {{Visualization and analysis of atomistic simulation data with OVITO-the
   Open Visualization Tool}},
Journal = {{MODELLING AND SIMULATION IN MATERIALS SCIENCE AND ENGINEERING}},
Year = {{2010}},
Volume = {{18}},
Number = {{1}},
Month = {{JAN}},
DOI = {{10.1088/0965-0393/18/1/015012}},
Article-Number = {{015012}},
ISSN = {{0965-0393}},
EISSN = {{1361-651X}},
ResearcherID-Numbers = {{Stukowski, Alexander/G-9695-2017}},
ORCID-Numbers = {{Stukowski, Alexander/0000-0001-6750-3401}},
Unique-ID = {{ISI:000272791800012}},
}

@article{ECHEVERRIRESTREPO2023111978,
title = {ABC-FIRE: Accelerated Bias-Corrected Fast Inertial Relaxation Engine},
journal = {Computational Materials Science},
volume = {218},
pages = {111978},
year = {2023},
issn = {0927-0256},
doi = {https://doi.org/10.1016/j.commatsci.2022.111978},
url = {https://www.sciencedirect.com/science/article/pii/S0927025622006899},
author = {Sebastián {Echeverri Restrepo} and Predrag Andric}
}

@article{HenkelmanA,
  author    = {Graeme Henkelman and Hannes J{\'o}nsson},
  title     = {Improved tangent estimate in the nudged elastic band method for finding minimum energy paths and saddle points},
  journal   = {The Journal of Chemical Physics},
  year      = {2000},
  volume    = {113},
  number    = {22},
  pages     = {9978--9985},
  doi       = {10.1063/1.1323224},
  url       = {https://doi.org/10.1063/1.1323224}
}

@article{HenkelmanB,
  author    = {Graeme Henkelman and Blas P. Uberuaga and Hannes J{\'o}nsson},
  title     = {A climbing image nudged elastic band method for finding saddle points and minimum energy paths},
  journal   = {The Journal of Chemical Physics},
  year      = {2000},
  volume    = {113},
  number    = {22},
  pages     = {9901--9904},
  doi       = {10.1063/1.1329672},
  url       = {https://doi.org/10.1063/1.1329672}
}

@article{Larsen2020Revisiting,
  author    = {Peter M. Larsen},
  title     = {Revisiting the Common Neighbour Analysis and the Centrosymmetry Parameter},
  journal   = {arXiv preprint arXiv:2003.08879},
  year      = {2020},
  url       = {https://arxiv.org/abs/2003.08879},
  note      = {11 pages, 8 figures}
}

@article{Hirel2015Atomsk,
  author    = {Pierre Hirel},
  title     = {Atomsk: A tool for manipulating and converting atomic data files},
  journal   = {Computer Physics Communications},
  year      = {2015},
  volume    = {197},
  pages     = {212--219},
  doi       = {10.1016/j.cpc.2015.07.012},
  url       = {https://www.sciencedirect.com/science/article/pii/S0010465515002817}
}

@article{
doi:10.1073/pnas.2322962121,
author = {Killian Sheriff  and Yifan Cao  and Tess Smidt  and Rodrigo Freitas },
title = {Quantifying chemical short-range order in metallic alloys},
journal = {Proceedings of the National Academy of Sciences},
volume = {121},
number = {25},
pages = {e2322962121},
year = {2024},
doi = {10.1073/pnas.2322962121},
URL = {https://www.pnas.org/doi/abs/10.1073/pnas.2322962121},
eprint = {https://www.pnas.org/doi/pdf/10.1073/pnas.2322962121}}

@article{BARUFFI2023115536,
title = {Equilibrium versus non-equilibrium stacking fault widths in NiCoCr},
journal = {Scripta Materialia},
volume = {235},
pages = {115536},
year = {2023},
issn = {1359-6462},
doi = {https://doi.org/10.1016/j.scriptamat.2023.115536},
url = {https://www.sciencedirect.com/science/article/pii/S1359646223002609},
author = {C. Baruffi and M. Ghazisaeidi and D. Rodney and W.A. Curtin}
}

@article{XU2024115934,
title = {Accurate complex-stacking-fault Gibbs energy in Ni3Al at high temperatures},
journal = {Scripta Materialia},
volume = {242},
pages = {115934},
year = {2024},
issn = {1359-6462},
doi = {https://doi.org/10.1016/j.scriptamat.2023.115934},
url = {https://www.sciencedirect.com/science/article/pii/S1359646223006553},
author = {Xiang Xu and Xi Zhang and Andrei Ruban and Siegfried Schmauder and Blazej Grabowski}
}

@article{Ding2018,
  author  = {Ding, Jun and Yu, Qin and Asta, Mark and Ritchie, Robert O.},
  title   = {Tunable stacking fault energies by tailoring local chemical order in {CrCoNi} medium-entropy alloys},
  journal = {Proceedings of the National Academy of Sciences},
  volume  = {115},
  number  = {36},
  pages   = {8919--8924},
  year    = {2018},
  doi     = {10.1073/pnas.1808660115},
  url     = {https://www.pnas.org/doi/abs/10.1073/pnas.1808660115}
}

@article{liu2018stacking,
  title={Stacking fault energy of face-centered-cubic high entropy alloys},
  author={Liu, SF and Wu, Y and Wang, HT and He, JY and Liu, JB and Chen, CX and Liu, XJ and Wang, H and Lu, ZP},
  journal={Intermetallics},
  volume={93},
  pages={269--273},
  year={2018},
  publisher={Elsevier}
}

@article{zhang2013first,
  title={First-principles study of stacking fault energies in Mg-based binary alloys},
  author={Zhang, Jing and Dou, Yuchen and Liu, Guobao and Guo, Zhengxiao},
  journal={Computational materials science},
  volume={79},
  pages={564--569},
  year={2013},
  publisher={Elsevier}
}

@article{denteneer1991energetics,
  title={Energetics of point and planar defects in aluminium from first-principles calculations},
  author={Denteneer, PJH and Soler, JM},
  journal={Solid state communications},
  volume={78},
  number={10},
  pages={857--861},
  year={1991},
  publisher={Elsevier}
}

@article{denteneer1987stacking,
  title={Stacking-fault energies in semiconductors from first-principles calculations},
  author={Denteneer, PJH and Van Haeringen, W},
  journal={Journal of Physics C: Solid State Physics},
  volume={20},
  number={32},
  pages={L883--L887},
  year={1987}
}

@article{Ghasemi_2019,
  title = {Nudged Elastic Band Method for Solid-Solid Transition under Finite Deformation},
  author = {Ghasemi, Arman and Xiao, Penghao and Gao, Wei},
  year = 2019,
  month = aug,
  journal = {The Journal of Chemical Physics},
  volume = {151},
  number = {5},
  publisher = {AIP Publishing},
  issn = {1089-7690},
  doi = {10.1063/1.5113716}
}

@inproceedings{Boerner_2023, series={PEARC ’23}, title={ACCESS: Advancing Innovation: NSF’s Advanced Cyberinfrastructure Coordination Ecosystem: Services \&amp; Support}, url={http://dx.doi.org/10.1145/3569951.3597559}, DOI={10.1145/3569951.3597559}, booktitle={Practice and Experience in Advanced Research Computing}, publisher={ACM}, author={Boerner, Timothy J. and Deems, Stephen and Furlani, Thomas R. and Knuth, Shelley L. and Towns, John}, year={2023}, pages={173–176}, collection={PEARC ’23} }

\end{document}